\documentclass{ws-procs9x6}

\def\etal {{\it et al.}}

\begin{document}

\title{HOW TO TEST THE SME WITH SPACE MISSIONS?}

\author{A.\ HEES}

\address{Jet Propulsion Laboratory, California Institute of Technology\\
4800 Oak Grove Drive, Pasadena CA 91109, USA\\
E-mail: Aurelien.D.Hees@jpl.nasa.gov}

\author{B.\ LAMINE}

\address{Laboratoire Kastler Brossel, CNRS, ENS, UPMC, Campus Jussieu, F-75252 Paris\\
IRAP, CNRS, UPS, 14 avenue Edouard Belin, 31450 Toulouse, France}

\author{C.\ LE PONCIN-LAFITTE and P.\ WOLF}
\address{LNE-SYRTE, Observatoire de Paris, UPMC,\\
61 avenue de l'Observatoire, F-75014 Paris, France}

\begin{abstract}
In this communication, we focus on possibilities to constrain SME coefficients using Cassini and Messenger data. We present simulations of radioscience observables within the framework of the SME, identify the linear combinations of SME coefficients the observations depend on and determine the sensitivity of these measurements to the SME coefficients. We show that these datasets are very powerful for constraining SME coefficients.
\end{abstract}

\bodymatter

\section{Introduction}

Since the development of General Relativity (GR), the solar system has always been a very interesting laboratory to test gravitation theory and to constrain hypothetical alternative theories of gravity. Until today, mainly two formalisms have been widely used at solar system scales to test the gravitation theory: the parametrized post-newtonian (PPN) formalism and the fifth force search.

Within the PPN formalism, the metric is phenomenologically parametrized by 10 dimensionless parameters~\cite{will:1993fk} that can be constrained independently from any underlying fundamental theory. The current constraints on these PPN parameters are pretty good and can be found in Ref.\ \refcite{will:1993fk}. 

The fifth force formalism consists in searching for a modification of the Newton potential of the form of a Yukawa potential parametrized by a range of interaction and an intensity~\cite{fischbach:1999ly}. The area of the parameter space excluded by experiments can be found in Ref.\ \refcite{konopliv:2011dq}. It can be seen that very good constraints are available except for small and large interaction distances.

Even if the constraints on these two formalisms are currently very impressive, there are still theoretical motivations to improve them (for some examples, see Ref.\ \refcite{hees:2012fk}). Moreover, it is also very interesting to look for deviations from GR in other frameworks than the two used so far. In particular, a consideration of a hypothetical Lorentz violation in the gravitational sector naturally leads to a parametrized expansion at the level of the action~\cite{kostelecky:2004fk}.  The post-newtonian metric resulting from this formalism (known as the Standard-Model Extension (SME)) is parametrized by a symmetric traceless tensor $\bar s_{\mu\nu}$ and differs from the PPN metric~\cite{bailey:2006uq}. Until now, the only tracking data used to constrain these SME coefficients are the lunar laser ranging (LLR) data~\cite{battat:2007uq}.

In this communication, we show how spacecraft tracking data can be used to constrain SME gravity parameters. For this, we determine the incompressible signature produced by SME on tracking observations. The procedure and the software used to determine these signatures are presented in Ref.\ \refcite{hees:2012fk}. 

\section{Simulations of tracking observations in the SME}

We consider three realistic situations: a two year radioscience link between Earth and the Mercury system corresponding to Messenger data, a 32 day Doppler link between Earth and the Cassini spacecraft during its cruise between Jupiter and Saturn corresponding to the conjunction experiment~\cite{bertotti:2003uq}, and a 9 year radioscience link between Earth and the Saturn system corresponding to Cassini data. For these three situations, we determine the linear combinations of SME coefficients to which the observations are sensitive, the signatures produced by these coefficients on observations and the sensitivity of these observations to SME coefficients.

The radioscience (range and Doppler) measurements of Messenger depend on the 4 linear combinations of the 9 fundamental parameters $\bar s_{\mu\nu}$:
\begin{subequations}
	\begin{eqnarray}
		\bar s_A &=&\bar s_{XX}-0.72 \bar s_{YY} -0.28 \bar s_{ZZ}, \\
		&&\hskip -25pt\bar s_{TX}, \\
		\bar s_B &=& \bar s_{TY} + 0.53 \bar s_{TZ}, \\
		\bar s_C &=& \bar s_{XY}+ 2.954 \bar s_{XZ} -  0.26\bar s_{YZ}.
	\end{eqnarray}
\end{subequations}
The 32 days of Doppler data from the Cassini conjunction experiment depend only on two linear combinations given by
\begin{subequations}
	\begin{eqnarray}
		\bar s_D &=&\bar s_{XX}-0.84 \bar s_{YY} -0.16 \bar s_{ZZ} +9.45 \bar s_{XY} + 4.1\bar s_{XZ} -0.72 \bar s_{YZ},\\
		\bar s_E &=& \bar s_{TX}+3.69\bar s_{TY} + 1.55 \bar s_{TZ},
	\end{eqnarray}
\end{subequations}
while the 9 year Range and Doppler data coming from the Saturnian system depend on 
\begin{subequations}\label{eq:saturn}
	\begin{eqnarray}
		\bar s_F &=&\bar s_{XX}-0.83 \bar s_{YY} - 0.17 \bar s_{ZZ}  - 0.76 \bar s_{YZ}, \\
	&&\hskip -25pt\bar s_{TX},\\
		\bar s_G &=& \bar s_{TY} + 0.43 \bar s_{TZ},\\
		\bar s_H &=& \bar s_{XY}+0.56 \bar s_{XZ}.
	\end{eqnarray}
\end{subequations}
\begin{figure}[t]
\begin{center}
\psfig{file=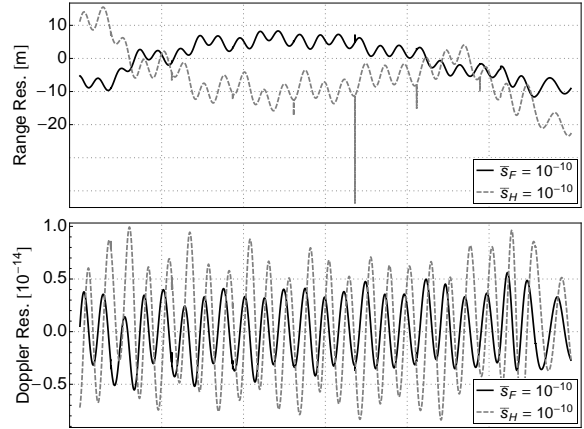,width=0.7\textwidth}
\end{center}
\caption{Incompressible signatures produced by some of the SME linear combinations (\ref{eq:saturn}) on range and Doppler observations of Cassini while orbiting in the saturnian system.}
\label{fig1}
\end{figure}

Figure \ref{fig1} represents the signature due to some of the SME linear combinations (\ref{eq:saturn}) on the Cassini radioscience measurements. These signatures correspond to residuals that would be obtained by a naive observer measuring data and analyzing them in GR (using standard procedure) while the correct gravitation theory is SME theory with the linear combinations taking the indicated values. The signatures are characteristic of the SME theory of gravity and should be searched for in the residuals of real data analysis. Similar signatures have been determined for the two other situations (Messenger and the Cassini conjunction) and for the other linear combinations but are not presented here. 

\def\tabsubcap#1#2{\noindent\centering\footnotesize(#1): #2\\\vspace{1mm}} 
\begin{table}[t]
\tbl{Estimated reachable uncertainties on SME coefficients}
{
\parbox[t]{0.3\linewidth}{\begin{center}\tabsubcap{a}{Messenger} \begin{tabular}{cc}
\toprule
Coeff. & Uncertainties\\\colrule
$\bar s_A$ & $1.1 \times 10^{-10}$\\
$\bar s_{TX}$ & $3.1 \times 10^{-8\phantom{0}}$ \\ 
$\bar s_B$ & $1.4 \times 10^{-8\phantom{0}}$ \\
$\bar s_C$ & $3.2 \times 10^{-11}$ \\\botrule
\end{tabular}\end{center}}
\parbox[t]{0.3\linewidth}{\begin{center}\tabsubcap{b}{Cassini conjunction}\begin{tabular}{cc}
\toprule
Coeff. & Uncertainties \\\colrule
$\bar s_D$ & $3.6 \times 10^{-7}$\\
$\bar s_E$ & $3.1 \times 10^{-3}$ \\\botrule 
\end{tabular}\end{center}}
\parbox[t]{0.3\linewidth}{\begin{center}\tabsubcap{c}{Cassini in orbit}\begin{tabular}{cc}
\toprule
Coeff. & Uncertainties \\\colrule
$\bar s_F$ & $8.6 \times 10^{-11}$\\
$\bar s_{TX}$ & $1.2 \times 10^{-8\phantom{0}}$ \\ 
$\bar s_G$ & $1.5 \times 10^{-8\phantom{0}}$ \\
$\bar s_H$ & $2.3 \times 10^{-11}$ \\\botrule 
\end{tabular}\end{center}} 
\label{tab}}
\end{table}

The comparison of the amplitude of these signatures with the accuracy of the measurements gives an estimate of the uncertainties on the SME coefficients that would be reachable in a real data analysis. The estimated uncertainties on SME coefficients reachable using Messenger and Cassini data are presented in Table \ref{tab}. One can see that the conjunction data are not interesting to constrain SME. On the other side, Messenger and Cassini data (while orbiting within the saturnian system) are very interesting and can improve the current LLR constraints on SME coefficients~\cite{battat:2007uq} by one order of magnitude. This gives a strong motivation to consider a test of SME using these datasets.

\section*{Acknowledgments}

The research described in this paper was partially carried out at the Jet Propulsion Laboratory, California Institute of Technology, under contract with the National Aeronautics and Space Administration \copyright \ 2013. All rights reserved. A.H. acknowledges support from the BAEF.


\begin{thebibliography}{x}

\bibitem{will:1993fk}
C.M.\ Will, {\it Theory and Experiment in Gravitational Physics}, Cambridge University Press, 1993; C.M.\ Will, Liv.\ Rev.\ Rel.\ {\bf 9}, (2006).

\bibitem{fischbach:1999ly}
E.\ Fischbach and C.L.\ Talmadge, {\it The Search for Non-Newtonian
  Gravity}, Aip-Press Series, Springer, 1999; 
E.G.\ Adelberger, J.H.\ Gundlach, B.R.\ Heckel, S.\ Hoedl, and
S.\ Schlamminger, Prog.\ Part.\ Nucl.\ Phys.\ {\bf 62}, 102 (2009).

\bibitem{konopliv:2011dq}
A.S.\ Konopliv \etal, Icarus {\bf 211}, 401 (2011).

\bibitem{hees:2012fk}
A.\ Hees \etal, Class.\ Quantum Grav.\ {\bf 29}, 235027 (2012).

\bibitem{kostelecky:2004fk}
V.A.\ Kosteleck\'y, Phys.\ Rev.\ D {\bf 69}, 105009 (2004).

\bibitem{bailey:2006uq}
Q.G.\ Bailey and V.A.\ Kosteleck\'y, Phys.\ Rev.\ D {\bf 74}, 045001 (2006).

\bibitem{battat:2007uq}
J.B.R.\ Battat, J.F.\ Chandler, and C.W.\ Stubbs, Phys.\ Rev.\ Lett.\ {\bf 99}, 241103 (2007).

\bibitem{bertotti:2003uq}
B.\ Bertotti, L.\ Iess, and P.\ Tortora, Nature {\bf 425}, 374 (2003).

\end{thebibliography}

\end{document}